# Rejecting Adaptive Interface

*Abstract*. Programs have to be designed in such a way as to make them looking good and being handy for all users. Adaptive interface, with all the numerous achievements throughout 30 years of its history, contains and in reality is based on one fundamental flaw – on the assumption that designer knows better than anyone else what is good for users in each and all cases. The new type of programs – *user-driven applications* – still deliver to users the best results of developers' work but at the same time give users the full control over applications and in this way really allows each user to change an application in such a way as he wants it to look at each particular moment. Users can move and resize each and all of the screen objects while an application is running, and this changes the whole programming philosophy.

## Introduction

Usually an introduction has to give a brief view of the main idea of the following article. I have to start this introduction with the words about what this article is **not** about; I have to write it in such a way because readers often interpret the very first words about movability of the screen objects in an absolutely wrong way. The whole thing described here is based on the movability of the screen objects, but the movability that is used in such a way and to such extent as none of other applications or systems used.

**First**, this is not about movie making or game design. In the movies everything is moving, but onlookers have no control over those movements; they can only watch. The games are not far away from the movies in this aspect. There is a scenario and whatever is done is simply one or another variant of this scenario. All movements of all objects are predetermined and already coded. All movements are predetermined by a designer.

**Second**. When people understand the previous remark and acknowledge that I am going to write only about such movements of objects that are controlled and performed by users and cannot be predetermined by developers, then comes a standard remark that there is also no novelty in it because in a famous Windows system everything can be moved by users; so looks like I am writing about the things which are known from the beginning of the World. Well, let us look at the moving of objects under Windows more closely. In this and similar operating systems the rectangles representing applications can be moved and resized. Also icons can be moved around the desktop, and that is all. These two types of objects can be moved on the level of operating system. Start any application and try to move any object inside. Is it possible? With an exception of few programs (like *Paint* and similar) nothing can be moved inside the programs.

The inner world of applications is the area about which I am going to write and this is the thing which is going to change the world of programming when all the inner objects of programs will be turned into movable and resizable.

Computers became such a big part of our life that nearly nothing is done without them. But computers are only the instruments which allow us to deal with more important objects – computer programs. These programs are now changing and even controlling the life of individuals and societies; by doing it those applications produce a phenomenon by itself. For the second time in history of mankind the virtual world became equally important if not more important than the real world around us. The first time it happened when religion was created, and there is a fundamental difference in our dealing with these two virtual worlds. People try to keep their religion untouchable throughout the time; any attempts to change the foundations of religion were and in many countries still are considered as the most serious crimes and are punished in the most severe way. The situation with the computer programs is opposite in every aspect: applications are constantly changed and the very existence of these virtual objects is due only to their everlasting changes.

Every program is an invisible link between two groups of people: on one side there is a single developer or a group of developers; on the other side there are users; their number can vary from single digit to several millions. If a program is really good, then users would prefer it not to be changed at all, but this is impossible, because this would be the end of any application. Also, the developers want to eat and their existence depends on the life of applications, which, like a bicycle, are alive only through movement. The developers are constantly changing their products and this looks like a classical process of evolution.

At the time when the term *personal computer* didn't even exist, the overwhelming majority of programs were scientific or engineering; a lot of them were developed for the projects initiated by the Department of Defense, but the tasks themselves and the computer programs to solve those tasks were of the mentioned types. The researchers who worked on those projects were mostly proficient in math; they easily learnt FORTRAN and turned their knowledge into algorithms and the code of working programs. The same person was a researcher, an author of the code, and the analyser of results; the output was mostly in numbers and could be easily changed by this person if he wanted to do it. The algorithms were huge; the programs had long listings and were often written in *spaghetti code*.[*] In a while you could see more specialization in programming area with different people being researchers and program developers, but everything else was the same. Researchers and program developers often worked in close collaboration and the problems of the output were solved immediately or after a short discussion on a personal level. The main thing is that at that time the developer of a program had a total control over the calculations and over the visualization. It was going this way for so long that became an axiom.

Things began to change rapidly with the beginning of PC era and in our days users of some programs can outnumber the developers by thousands or millions; the solving of interface problems through personal communication became inappropriate. It's impossible to imagine that thousands or millions of users would be absolutely satisfied with the designer's ideas of interface and the solution to this problem was obvious: give users an instrument to adapt the interface to their demands. For the last 30 years the design of interfaces for computer programs was influenced mostly by the ideas of adaptive interface. The computer revolution tremendously increased the number of developers, so the stream of different ideas for adaptive interface was enormous. Not all of these ideas became popular but some of them are; for example, dynamic layout promoted now as the mainstream for today and for the nearest future.

In reality adaptive interface gives an illusion of users' freedom of solutions; it never gives user an opportunity to do whatever he personally wants, but only allows to select among choices which were previously considered by developers as good enough for one or another situation. This is a fundamental flaw of adaptive interface. If it happens so that some user shares the developer's ideas about the best looking interface, then this user is really lucky and can get exactly what he wants. But it is a very rare situation. More often than not you either hate the interface of a program or strongly dislike it but in any case you have to work with it because you are more interested in the results that you can obtain from this program. Adaptive interface is always called a friendly interface regardless of whether you like the introduced example of it or strongly dislike. It looks like a mockery to me.

Dynamic layout demonstrates the worst type of developers' – users' relation; it doesn't even allow any choice but simply enforces the developer's view of the good looking program. "You have to like whatever we are giving you" – this must be the motto for developers using dynamic layout. Suppose that you have a resizable form (dialogue) with a single control inside, then with the resizing of the form this control changes its size so as to fill the same percent of the form's area or keep the constant spaces between the control and the borders of the form; this is an idea of the classical anchoring. It is very easy to implement and the results will satisfy nearly everyone, though I mentioned two possibilities even for this extremely primitive case. But what about the case with two controls inside? They can resize both, or only the first one, or only the second. And the programmer has two choices: either to put into code the variant he personally prefers or to add an instrument which allows users to select one of three variants. For the case of two controls you get three possible variants. What if you have not two but three, four, or more objects inside? What are you going to do in case of 30 or 40 objects? Do you think that you would be capable to organize a selection between all the possible variants and would like to introduce the users to such a system? I doubt. The only solution is to implement some variant and explain to users that this is the best variant which they would like. That is how all the popular (and less popular) programs are designed. "You have to like whatever we are giving you". In general there is no explanation to users and they simply have to use whatever they are given. Only because they are allowed to change the size of a form and on this action the implemented dynamic layout changes the sizes and positions of the inner elements such interface is called user-friendly!

The good developers are well aware of those interface problems and try to solve them. Unfortunately, they continue to look for a new solution in the standard way: if users are not satisfied with some part of a big program, then this part is redesigned in such a way as to give users more flexibility in this segment. Simply a well known or new and more sophisticated version of adaptive interface is used. It's exactly like trying to develop a real perpetual motion machine by using new materials which were not available years or decades ago. One more effort and we'll get it! In 1775 the Royal Academy of Sciences in Paris issued the statement that the Academy "will no longer accept or deal with proposals concerning perpetual motion". In which year to come the developers will understand that no version of adaptive interface can produce an interface which is excellent for each particular user? Adaptive interface is a popular highway to a dead end.

Does it mean that we'll never give users anything better than adaptive interface? Is there any other way of changing programs? Thoughts about this alternative way and work on realization of the ideas helped me to produce the applications of absolutely different type. I call them *user-driven applications*; if you read this article and look at the mentioned programs you will understand why I gave such a name to the new type of programs.

---

[*] Don't mix this term with *spaghetti westerns* which also flourished at that time. Dijkstra's letter to the Editor of Communications of the ACM, published in March 1968, marks the beginning of structured programming. I doubt that anyone in his clear mind would like to return to the type of programs that were popular before that, while a lot of people (and I am among them) would like to find a spare time, sit down, and admire once more *The Good, the Bad and the Ugly* (Sergio Leone, 1966).

## What is so new in user-driven applications?

Let us look at the interface problems from a different point; from the point which is far away from our current time mark. Imagine that we are somewhere 50 or 100 years ahead in time and try to solve some problems on what will be called a computer at that time. Whether we are going to sit at the desk or stay in front of a wall, these are unimportant details which we can't predict. More general things are much more important. There must be some information screen with the objects showing us the needed information in different ways. We will have to communicate with computer in one way or another and there are not too many choices in organizing it. We – people – will have the same mouse, eyes, ears, and hands; these are our only options for communication with computers. It really doesn't matter what will be our preferable way of communication at one or another moment: voice commands, rolling of our eyeballs from side to side, some descendant of a mouse or stylus, or there will be some new device more powerful and efficient for our hands. The main thing is that in one way or another we will be still capable to communicate with computer and order it to get the new data for our tasks under solving and to change the view, position, and sizes of all the screen objects through which we get the results. This is the most important and fundamental feature: we – users – have to have a very powerful and efficient way of controlling all the screen objects. Nobody else but each user personally will control every visualization parameter of the received information.

The information itself is provided by an application. You can call a core device CPU, engine, solver, or by any other name. It gets the input data, works on it according to some algorithms, and returns the results. The outcome depends only on a set of input data and determined algorithm. This outcome does not depend at all on the way it is going to be shown, but the interpretation of the results by the users can significantly rely on the view of the results and the preferences of each user.

The fundamental interface problem is in the fact that up till now interface is always run by developer! Nobody before declared that it can be organized in a different way and this became an axiom. It always was, it works this way in all currently used programs (except mine!), and it always will be! Are you sure? It's not a nature law like a law of universal gravitation. The developers' control over interface is a historical fact and nothing more. It is based only on the currently used programming technique, but what if you change this basis? **Can you take the control over interface from developer?**

"In science, finding the right formulation of a problem is often the key to solving it…" [1]. Let's formulate the interface problem in its shortest way without any secondary details; maybe this will show a way to solution.

   a. Developer writes the code for an algorithm or uses some method from a standard library.

   b. The results must be shown to users.

   c. Developer provides the visualization (this is a big part of interface) in the best way he can.

   d. Many users do not like the output view and complain.

   e. Developer provides several variants of output; users may select one of them (adaptive interface).

   f. Users still do not like the output view. Return to the previous line with more (or different) variants.

For the last 30 years developers and users are enclosed into an infinitive loop (e – f) with no way out. Developers continue to look for the solution in this infinitive loop. Do you see the way out?

The solution to fundamental interface problem can be only a drastic one: exclude developer from the infinitive loop (e – f) and get rid of this loop by giving users the full control over visualization. It means not only the control over auxiliary parameters like font and color but over the main things: places, sizes, and appearance of all the screen elements. To implement such thing, all the screen objects must be turned into movable and the full control over their movability, resizability, and all possible changes must be given to users. Full control means that developer provides a default view of an application and turns the control to users, but there is no second loop control from developer over the users' doings. This change in design and use of programs can be done on two main conditions.

- An algorithm to turn any object into movable / resizable must be easy to implement for all objects without exceptions.

- Moving and resizing must be easy and similar in use for all objects.

There were works before which tried to implement the movability of the screen elements <u>by users</u>. I have a high respect to those who were thinking about such algorithm years ago. I have already designed some programs with the movable elements 20 and 15 years ago; I am familiar with other applications in which some movability was introduced, but there is a fundamental difference between previous attempts (both mine and of other people) and what I demonstrate now. The movability in all those old systems and programs was limited to some very limited set of objects and never spread to wider area of elements. 20 years ago I needed the movability of some objects in my scientific programs and developed it exactly for those objects, but that old algorithm could not be used for an arbitrary type of objects. My old algorithm was specialized and I saw the same limitation in all other programs where I met with movability of the screen objects. <u>I never heard or read about an algorithm that allowed to turn any object into movable and I never heard about a system (application) in which</u> **<u>everything would be movable</u> by user**. Sorry to mention, but I have several reviews on my work with the statements like this: "This is not new. It was already done in 1980-s and 1990-s by this and that person…" Such review never comes with the exact link on a scientific paper but only with the names. And I spend again and again a lot of time on looking through the old papers only to find once more that there was nothing of the mentioned type; if anything was done then it was done with a lot of limitations. More often than not all those old papers are about the <u>movability of objects by developers</u>! Those old researchers didn't try to make false statements and I even ran into an article in which the very first sentence of an abstract declared that "proposed technique can be used **by developers** to move objects…", but… Reviewers are mostly respected people, but they consider themselves too busy to go into such small details. There were some words about movability in the old paper from the previous century and they are sure that everything was already done before. If you also think that everything about movability of the screen elements was thought out and published 15 years ago, then you better save your time and stop reading this article. Others may find very interesting things in the further text, but it would be much better if they look at the program in parallel with reading. Don't trust me simply by the words; try yourself.

Maybe one of the well known previous attempts related to movability was done in the Morphic interface construction environment [2]. It was a good system at its time and some version still exists, but it has a lot of limitations and looks so archaic in comparison with what I have designed that I don't want to go into detailed comparison. I have written a book about my algorithm and its implementation [3]; there is a whole theory of user-driven applications, so it is definitely not a short text. From the same site you can also download shorter texts which will give you a quicker understanding of the new ideas. To understand that there is no comparison between the Morphic proposal from 1995 and what I demonstrate now in my applications you have to start an application which accompanies my book and look at some examples there. Any examples by your choice; there are a lot of them. Unfortunately, I am well familiar with the situation when people do not bother themselves with looking at the working application and write their conclusion without ever seeing any user-driven applications. Looks like a very clever British scientist William Gilbert wrote [4] exactly for such "specialists": "In the discovery of secret things, and in the investigation of hidden causes, stronger reasons are obtained from sure experiments and demonstrated arguments than from probable conjectures and the opinions of philosophical speculators."

The main idea of user-driven application can be formulated in one fundamental law: user can do anything. User can move, resize, and change any screen element in any way he wants. To receive such a result, something must be done with the screen objects. When I began to work on the algorithm to turn screen objects into movable, there were two obvious requirements: it has to work without problems with any objects and it must be simple. What is really interesting in the proposed algorithm is its tendency for simplicity: for example, the mechanism of declaring movable and resizable a complex plotting area with an unlimited number of scales and related comments is nearly as simple as for a simple rectangle. Even if some of the following examples look like a science fiction to you, you must be aware from the very beginning that all of them are taken from the working application which is available at www.sourceforge.net in the project MoveableGraphics. The application is available together with all its codes!

There was a period in the history of programming when code for calculations was intermixed with code for visualization and this was considered normal; later two parts of code were separated and the benefits were high. Up till now both parts continue to be under developers' jurisdiction; it's the time to separate also control over them. The idea of taking the control over visualization from developers and giving it to users looks absolutely heretical to nearly all developers. The prevailing view among developers about users of their programs is short and never publicly expressed "Users are idiots". I heard it not once in private professional discussions and every time I laughed because I made simple extrapolation and then the developers in Microsoft must express the same opinion about all those using Visual Studio. (An excellent analogue of a feed chain.)

The structure of nearly any program can be represented by three main parts: input data preparation, calculations on this data, and the demonstration of results. There are some programs which look like lacking the middle part of this trilogy (take some data from a database and simply show it), but I'll show further on that even such applications fit very well with the ideas of user-driven applications. Now take a list of paper and write down who in the pair *developer - user* is responsible for each of three stages.

The next table shows the responsibility for three stages of a program. In the <u>currently used applications</u> users are responsible only for providing initial data, while two other parts are controlled by the designer of a program. Certainly, in the real applications the responsibility is not divided as strictly as it is shown in this table. For example, a developer provides some methods to check the input data; if you consider this checking as some part of the first column ("Input data") then a developer has some control over that column. When some form of adaptive interface is used in a program, then users have some influence over the output. But as I explained before, this users' control over the output is organized according to the developers' ideas and only inside the boundaries of what is allowed by designer of a program. Thus, in general, user is responsible for the input data, while everything else is ruled by developer.

|  | Input data | Calculations | Output (results view) |
|---|---|---|---|
|  | R e s p o n s i b i l i t y | | |
| Currently used programs | User | Developer | Developer |
| User-driven applications | User | Developer | **User** |

**Table 1**. Responsibility for the stages of the currently used programs and user-driven applications

Looking at the two bottom lines of the table you can see that from the point of applied comparison the <u>user-driven applications</u> have one major difference: in such programs user controls the output view. You can immediately ask a question of whether such switch is really so important. The answer is YES, ABSOLUTELY. Throughout the last five years I have designed a lot of user-driven applications. Those programs vary from the simplest examples to demonstrate some features of my theory to the most complicated applications from different areas; for example, it can be a scientific program for the area of thermodynamics or a huge program to organize and work with a personal photo archive. I am absolutely convinced that a switch from current day programs to user-driven applications means for users more than an old switch from DOS to Windows. The book [3] is accompanied by a big demonstration program with more than 100 examples; you can look through this program and see yourself that I am right in my statement.

This outstanding result after passing the control over the output from developer to user is due not only to the obvious switch in score between users and developers from 1:2 to 2:1 (if the score is calculated by the responsibility over programs' parts as shown in **table 1**), but due to the way we work with applications. The table shows the responsibility by stages of applications, but in time the real process is slightly different. The programs that we use are mostly interactive; users constantly check the output and, based on this analysis, they constantly change not only the view of the results but also the input data and start the calculations again. Thus users keep the control over running application.

Let's take for analysis a very simple task: suppose that we need a program to calculate the square of an area limited by function *f* on interval [a, b]. If *f* is a real-valued function that admits an antiderivative **F** on a closed interval [a, b], then the needed square is calculated according to the equation from **figure 1**. This formula is called a [second] fundamental theorem of calculus; it is known for around 350 years and is widely used in a lot of engineering and scientific tasks. Isaac Barrow proved a

$$\int_a^b f(x)\,dx = F(b) - F(a)$$

**Fig.1** Fundamental theorem of calculus

generalized version of this theorem and then published his *Geometrical Lectures* in 1669. As you see, this theorem was used long before the invention of computers and eventually found its way into the code of nearly every engineering program.

The underlying math is still the same in a lot of our day tasks, but everything else is absolutely different. You can find similar examples in [3], so I am going to describe the realization of such a task in case of user-driven application.

- Users need to calculate an area, so an area is drawn on the screen.

- This painted area can be moved around the screen to any location; just press it with a mouse and move.

- The size of an area can be changed by moving the left (parameter *a* in the original equation) or the right (parameter *b*) border. This change of an interval is done in the same easy way: press the border, move it, and release at the new location.

- The function *f* can be changed in different ways. You can select one line from the predefined list of functions, or you can type in the text of a function using the standard mathematical notation and the interpreter will do his work, or the upper border of the area can be shown as a polyline with sensitive dots at the joints of the neighbouring segments. The number of segments can be easily changed by adding and deleting joints; all the joints are movable, so this is one more way to change the function.

- Not only the sizes but also all the colors of the picture can be easily changed by users.

- For any piece of textual information on the screen the color, font, and position can be easily changed by user at any moment.

Examples from the book make it more vivid, but while looking through this list of available changes you can see that <u>everything can be changed by user</u>. The only piece left in the program, which is still under the developer's control, is the implementation of math equation somewhere deep inside the code. In nearly all the programs the real calculation is passed to some standard libraries. But whatever is done with the program throughout the time of its running is done only by user. User gets the whole control of everything; this is the reason to call such applications *user-driven*.

## Rules of user-driven applications

The development of such programs is done according to several basic rules. The rules were not thought out prior to the development of user-driven applications but were formulated as a result of practical design of many programs. These are not the artificial rules that somebody (me!) advises you to obey. When you start developing applications of the new type you quickly find that these rules are absolutely logical and you will implement them in all the cases without any exceptions; in a short period of time you will obey them absolutely automatically. In the book I wrote in details about each rule and explained why they must be meticulously carried out. Here I want only to mention those rules of user-driven applications.

- All the elements are movable.
- All the parameters of visibility must be easily controlled by users.
- The users' commands on moving / resizing of objects or on changing the visibility parameters must be implemented exactly as they are; no additions or expanded interpretation by developer are allowed.
- All the parameters must be saved and restored.
- The above mentioned rules must be implemented at all the levels beginning from the main form and down to the farthest corners.

No one is born with the knowledge of the surrounding world and no one has an inherited knowledge of how to work on computer and deal with the screen objects. When you read or hear that you can deal with some type of objects in a natural way it means that you can work with them in exactly the same way as you were dealing with similar objects on previous occasions. For example, if user needs to resize an object, then such person would try to grab the border of an object and move it in the needed direction. For simple objects of a primitive shape, like straight lines, circles, or rings everyone will try to do the resizing in the same way and the result will be exactly as expected. For more complicated figures and for the objects with some details the "natural" feeling of the developer can differ from "natural" feeling of some users; this can cause a misunderstanding and create the users' anger against the "stupidity of those developers". What will you say about the resizing of a polygon with an arbitrary positioning of its apices? Or what do you expect on the natural (by borders) resizing of a picture of apartment building? Do you expect the number of storeys and windows on each floor to change, or keep their numbers unchanged but change the size of each window, or keep the same number and size of windows but change spaces between them? I would call it "natural according to the rules of particular application", but the less you use the special rules the easier for users to deal with an application.

The most natural way of moving any object is to grab it by any point and move. Surprisingly, we have an exception from this absolutely natural rule in the case of the most obvious and well known example – the moving of windows in the Windows system. It is organized in such a way that you can't move those windows by any inner point but only by the special title bar. This is so unnatural that no one would even think about such possibility until he is told about it. In reality, moving of windows by any inner point is easily organized by adding two – three lines of code; I show this in my applications, and I am not the only programmer to demonstrate such a thing.

The first and the most important rule of user-driven applications declares the movability of all the screen objects. Movability has different forms as there are forward movement, resizing, reconfiguring, and rotation. We have a wide variety of screen objects which differ in shape and complexity. Nobody knows what peripheral devices will be the most popular 20, 50, or 100 years from today. With my algorithm and applications I demonstrate that movability and resizability of all the screen objects changes the way in which we deal with computers and programs; for these demonstrations I use an instrument which looks to me the most appropriate at the current moment – a mouse. I try to demonstrate these revolutionary changes with the examples from different areas and with a lot of different objects. I also try to minimize the number of rules under which all these examples work, but I need to implement several of them in order to make the use of the new applications as easy as possible. The following rules are not mandatory, but they work in all of my applications to minimize the efforts of any user in dealing with any program of the new type.

- Everything is moved and resized only by mouse. Nothing else is needed.
- Everything is moved and resized in the "press – move – release" way.
- All graphical objects are moved by any inner point and resized by borders.
- Controls are moved and resized by different parts of their frame.
- Forward movement, resizing, and reconfiguring of any object are done by pressing the left button; rotation is done by the right button.

## Let's begin with graphical objects

The next table shows in a short form how these rules are applied to the graphical objects of the most popular shapes.

| | |
|---|---|
| 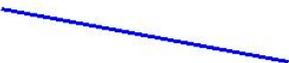 | Two end points of a line can be moved to an arbitrary position on the screen. By pressing any inner point a line can be moved around the screen or rotated around its central point. |
| 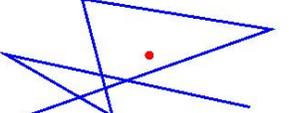 | The end points of all segments of a polyline can be moved independently. By inner points of segments the whole figure can be moved and rotated. The center of rotation can be placed anywhere; in my example the center of rotation is also a movable point. |
| 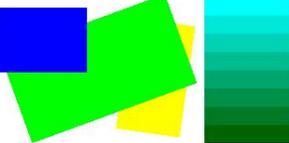 | Rectangles are the most widely used among the screen objects; while all of them can be moved by any inner point, there are many variants of their resizing. It is always done in the most natural way by border, but the moving of any side can be either independent or have some side effect. For example, the ratio of two sizes can be fixed, or two opposite sides can move symmetrically. Rotation usually goes around the central point but can be easily organized around different point. Rectangles may have inner movable partitions. |
| 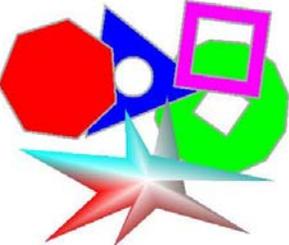 | Rectangles form a subset of polygons, but because there are many different variants for resizing rectangles and many types of other polygons, then for the purpose of better explanation I separated them in this table.<br><br>Polygons can be regular or arbitrary by shape. They can be solid or with the inner holes; the shape of a hole can be identical to the outer border of a polygon or can be different. Some polygons allow any transformations; others always stay convex. Minimum allowed size of a figure can be set to prevent its accidental disappearance during the resizing while in other cases such squeezing can be used to delete an object. |
| 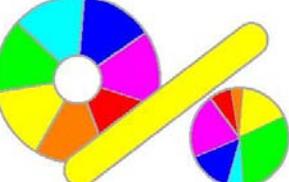 | Circles, rings, strips with the rounded ends, and some other figures not shown here can be united into one group because similar technique is used for resizing of objects with the curved borders. Circles and rings can be unicolored or multicolored; in the last case the partitions between segments can be made movable. For circles and rings there is an obvious center of rotation; for other objects it can be decided in one way or another. |
| 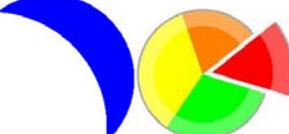 | Crescent can be resized by two horns and two opposite points of its widest part. Rotation can be done by any inner point, but the center of rotation is not obvious. A sliced pie can be moved and rotated as a whole object regardless of whether there are slices staying apart or not; a whole pie can be resized, but also each slice can be zoomed individually. These are examples of objects with no "natural" places to start some of the movements. |

**Table 2**. Different shapes of screen elements

As you can see from the shown examples, they differ very much from each other in shape. Yet, the algorithm to turn them all into movable and resizable is the same and very simple. There are some often used steps and the above shown objects can be used to learn the whole technique, but for each type of objects there can be more then one solution and developer can decide about using one or another. There is an exact analogue with integral calculus. You need to understand the idea of integration and to know several often used expressions and methods; with this knowledge you can use integrals to solve a lot of problems about which the inventors of integrals had no idea at all. To design user-driven applications, you have to understand the basic things about turning objects into movable / resizable and to be familiar with the demonstrated examples of the most popular shapes; on this basis you can think out an easy enough way to turn any object into movable / resizable.

For example, I decided to write a simple application for kids; this is one of the examples from [3]. It is called *Village*, because users can design a village of their own. **Figure 2** shows the types of houses which are used to construct a village. Looking at these buildings, you can easily find the similarities in shape with some graphical primitives from the previous table. Houses of four different types have a rectangular main part, while their roof is a polygon; the fifth type of buildings has a shape of semicircle. Similarities in shape with the graphical primitives also mean the same "natural" commands for resizing: all rectangular parts are resized by sides and corners; polygonal roofs are reconfigured by moving apices; semicircular hangar can be resized by any point of its roof. Thus all the buildings are resized by the borders and moved by any inner point. User of a program (an artist in front of the computer) decides about the number of needed buildings, their types, sizes, colors, and positions on the screen. At any moment the whole image of a village can be saved in order to continue the construction some time later. There are also some other very helpful commands which are nearly standard for every user-driven application. One of the main advantages of the proposed algorithm is its simplicity; the code does not depend at all on the number of movable objects or variety of their types.

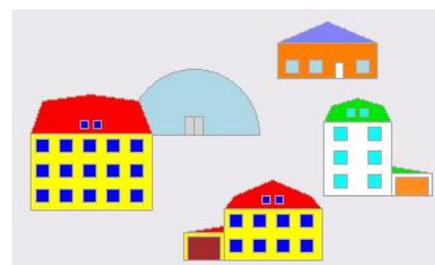

**Fig.2** Buildings for a village

## A small community of controls

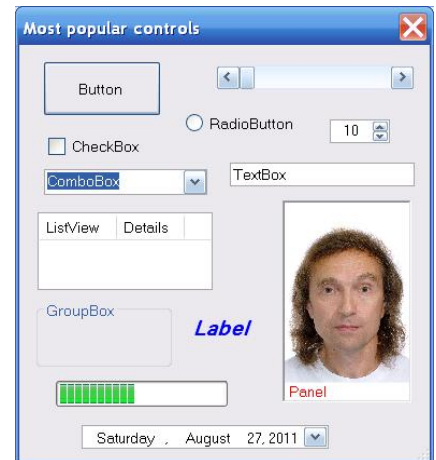

With all their infinitive diversity graphical objects constitute only one big group of screen elements; another and very different group consists of controls. The wide spread of predefined controls made programs much easier both in development and use. Designers have a wide enough set of construction blocks – between 20 and 30 most often used controls – to develop programs faster and in more reliable way. Users see the familiar set of elements in all the programs and the behaviour of these elements is always the same in all the applications; if you met a control once in some application, then you know exactly how to deal with it in other programs. **Figure 3** shows the view of the most popular controls which can be found in the majority of applications. To prepare this figure for current publication, I have to write a small application containing a set of controls. I place controls in such a way which looks good for me, compile and run a program, grab its view from the screen and save an image in a file. What have I to do if on the next day of writing this article I do not like the overall view of this picture and decide to change the sizes and positions of some elements? There are two different solutions: a standard approach and the new one.

**Fig.3** Popular controls

Standard approach means that in order to change the view of a form I go into Visual Studio, move and resize the controls in such a way which looks good at this particular moment, and then repeat all further steps starting with compilation.

But I know about the advantages of user-driven applications; now all my programs are designed only in this new way even if an application is going nowhere out and I am going to be its only user.

The first step in the development of analogous but user-driven application is exactly the same: use Visual Studio to place all the controls of the form in the best possible way. New application is not supposed to be sent to users with an instruction "Assemble it yourself"; developer puts the same efforts and all his skills into best possible design as he was doing for years, so the default view of this version is exactly the same as for a previous one. But all the controls in the application of the new type are declared resizable (by setting the corresponding values for two of their properties) and there is a small addition in the code as all elements are going to be moved and resized with a mouse. Because all the movements are done only by mouse then **MouseDown**, **MouseMove**, and **MouseUp** events are used.

When I started to write this article, I didn't want to include any code into the text as everything is explained in detailed way in the book. Later I decided to show some code to prove my thesis about the easiness of using the algorithm; now I have to write several words about it. There is an object of the `Mover` class; all the controls are registered with this *mover* object and after it this *mover* supervises the whole moving / resizing for all elements registered in its queue. Code for each of three used mouse events includes a call to one or another method of the `Mover` class. That is all that is needed to organize moving / resizing in a majority of situations; further details are in the book.

```
Mover mover;
mover = new Mover (this);
private void OnLoad (object sender, EventArgs e)
{
    mover .Clear ();
    foreach (Control ctrl in Controls)
    {
        mover .Add (ctrl);
    }
}
private void OnMouseDown (object sender, MouseEventArgs e)
{
    mover .Catch (e .Location, e .Button);
}
```

```
    private void OnMouseUp (object sender, MouseEventArgs e)
    {
        mover .Release ();
    }
    private void OnMouseMove (object sender, MouseEventArgs e)
    {
        if (mover .Move (e .Location))
        {
            Update ();
            Invalidate ();
        }
    }
```

Does it look like too much work in order to obtain an unlimited flexibility with all the elements in view? Now, if I want to rearrange this form (application, picture), I don't need to go into Visual Studio. Instead, I move and resize the controls in the running application and set the view I need.

This is a simplest type of example – only several solitary controls in a form and nothing else – but this example perfectly demonstrates the difference between standard programs and user-driven applications. If I am a user of a standard program then I have no chances to change its view according to my preferences; if dynamic layout or any other type of adaptive interface is implemented in the program, then I have a chance to make some changes, but all those changes are predetermined by developer. With the user-driven application the situation is different. A developer provides the movability and resizability of all elements; this gives user the full control over the view of an application; a developer is not controlling users in all their doings.

The technique for moving and resizing controls slightly differs from the technique used with graphical objects. "Thanks" to the origin of controls, I can't simply grab them by inner points; both moving and resizing of controls are done by the areas of their borders. (For more detailed explanation look into [3].)

## A stream of ideas for groups

Solitary screen objects play a significant role in many applications, but maybe more important are groups of elements. In complex programs with a lot of objects on the screen, designer has to inform users that, for example, several controls are used to change a group of closely related parameters or are used to solve some subtask. For the last 15 years developers have only two options to show any group on the screen: either to use `GroupBox` or `Panel`. Both are controls themselves; you can see them among the solitary controls at **figure 3**. Controls of these two types have perfectly visible borders and allow to place other controls inside those borders; panels also allow painting inside their area. The switch to movable elements allows to turn `GroupBoxes` and `Panels` into movable / resizable as any other controls. Much more important that the new algorithm allows to design absolutely new groups which are simply unthinkable in the standard programs. Looks like there is no end of new ideas in this area; the next table shows only several possibilities.

| | |
|---|---|
| 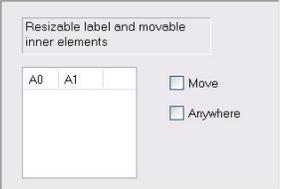 | This is a standard panel turned into movable and resizable. The panel is populated with controls which are also movable / resizable. As was explained earlier, any control can be moved and resized only by the areas of its border. It is a well known practice (thus it is natural) that rectangles can be resized by corners and small areas in the middle of its sides; all other parts of the control's frame are used for moving. |
| 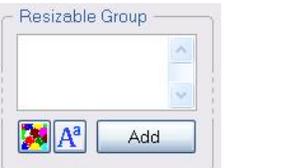 | `Group` class can be used by the admirers of dynamic layout. Such group can be moved by any inner point and resized by the frame. The changes of the inner elements on resizing the frame must be predetermined by designer; this is absolutely against the rules of user-driven applications, so I never use this class in my design; the class was designed only for the purpose of demonstration. |
| 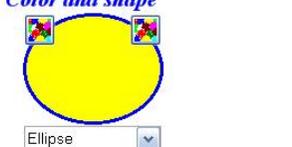 | This type of groups is used when a configuration of the group is fixed and not going to change, so there is no resizing. The whole group of elements can be moved around the screen by any inner point. The shown group has no frame, but it can be added. |
| 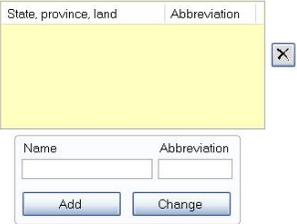 | Only one example from many variations on the "dominant + subordinates" theme. When a dominant element (in this case a colored `ListView` control) is moved or resized, then all subordinates (in this case a small button and a group) retain their relative positions to the dominant element. Any subordinate can be moved and resized independently of all other elements. In the book I demonstrate examples with different sets of classes allowed to be used as dominant and subordinate elements. |
| 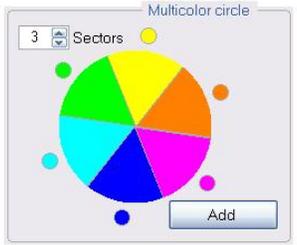 | Though this group looks similar to ordinary `GroupBox` but it is definitely not. The minor differences are the existence of graphical objects inside the group and the possibility of moving the title and putting it anywhere between the left and right sides of the frame (take it on trust or better try the program). The main differences are the movability of the group by any inner point and the elasticity of the frame. Inner elements can be moved and resized; the frame automatically adjusts to the united area of all inner elements. |

**Table 3**. Several variants of groups

Two things mentioned in the description of the last group are the main features of the `ElasticGroup` class which became the main element of design in many of my applications. The first thing – moving by any inner point – makes the moving of such groups as easy as moving of any graphical object. The second thing – adjustment of the frame to the united area of inner elements – is the result of very important idea implemented into design, so I want to write about it.

One of the best known forms of adaptive interface is the dynamic layout which is promoted by Microsoft as the best practice of design for today and for the nearest future. Implementation of dynamic layout means that designer decides about the new places and sizes of all inner elements whenever user resizes an area (usually it is a whole form, but it can be a panel with elements). I strongly oppose the full developer's control over the view of applications and I think that dynamic layout can be used and really useful only in case of a single element inside when this element has to occupy maximum available space. Situation with such a single element is very rare; in all other cases the dynamic layout has no right to exist.

If several elements are united into a group, then the frame of this group is definitely not the main element, but those inner elements are. It's not a problem at all to design a group moved by any inner point but using the ideas of dynamic layout; there is a `Group` class which works in such a way. I designed this class for the purposes of demonstration and comparison, but I am not going to use this class in any real development.

There is no place for designer's control over the view of user-driven applications, so there is no chance for inner elements to be positioned and resized according to some designer's ideas. User can organize elements inside an `ElasticGroup` object in any way he wants; the frame adjusts its position to all those changes inside only to make more obvious the area of the group. Frame of such group plays only an auxiliary (informative) part. The `ElasticGroup` class is used in many of my examples; maybe the best of them to show here is an application to deal with personal information. I didn't link it with any real database, so you are not going to see any data inside the inner elements, but all the possibilities of rearranging the view are demonstrated.

**Figure 4** shows some view of an application in which personal information from database can be viewed and edited. Objects of the `ElasticGroup` class can be easily used as nested elements. Database contains a lot of information about each person; in my example I structured this information into five obvious inner groups plus there are several controls with and without additional comments; these controls show the information not included into any group.

The `ElasticGroup` class itself and the demonstration programs on the basis of this class are designed according to all the rules of user-driven applications. All elements can be moved and resized. For some people it would be a question of their personal preferences whether to put a comment to an associated `TextBox` control on the left, on the right, or above. But the same easy moving of elements allows to solve in seconds much more important problem. There are countries in which addresses are shown in the order you can see inside the *Address* group; there are other countries in which the order is different and starts with the country name; the view of the group can be rearranged by user at any moment. You don't see the `TextBox` for country at this figure because the corresponding pair of elements was hidden by user; if you work only with the local staff then you don't need any information about country (and maybe Province also) on the screen. Any elements can be hidden or restored in view at any moment; groups can be hidden and restored in view in the same easy way. Color and font for any element can be changed individually; color and font for any group can be also changed. If there are nested groups, then there is a choice of either spreading these new visualization parameters on the inner groups or not. The rules of user-driven applications declare **entire** users' control over visibility parameters, so spaces between the inner elements of a group and its frame can be changed, the frame itself can be hidden or shown, the title of a group, if shown, can be moved between left and right sides of a group and this can be done regardless of the frame visibility. There are other things which are discussed in the book but which I am not going to discuss here. Full control really means full control! On closing this example, the currently used parameters of all the elements are saved in `Registry` (or in a file); the next time you start this application you see exactly the same view; none of your previous efforts on changing an application would be lost.

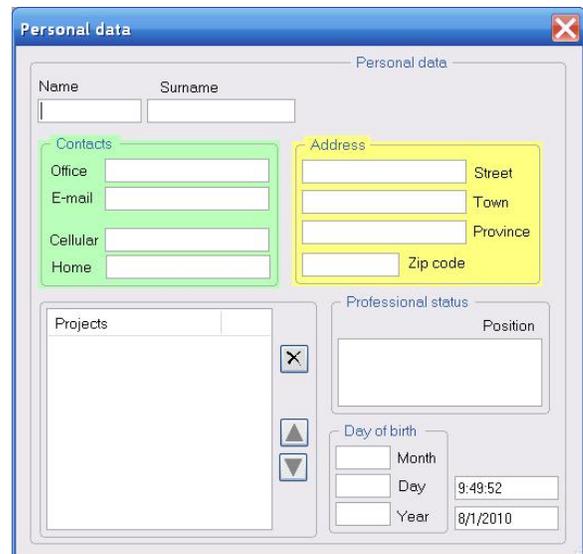

**Fig.4**  Some view of PersonalData application

I hope that every reader of this article clearly understands that there is no correlation at all between the work of this application on exchange of information with a database and a view of application. For a real application of such type the designer has to add two buttons to read and save information and to provide the correct reading and writing of data. The developer is responsible for correct work of an engine! In **Table 1** this stage of an application was called Calculations. There are no calculations in this application; instead there is an exchange of data with database; this is the only stage which is absolutely controlled by developer. The work of engine in any program is always hidden from view and the developer has the full responsibility over that hidden part. Whatever is visible is used either to enter data or to view the results; both stages are under the full user's control.

## Complex user-driven applications

As it happened many times in the past, some idea, which is born to solve problems in one area, turns out to be very valuable not only for this particular area but for many others also. The idea of total movability was born in the area of scientific applications which is my favorite area for decades. Throughout the years of my work as a scientist and a programmer I designed big and very complicated applications for different areas (applied optimization, voice analysis and speech recognition, telecommunication, electricity networks, thermodynamics) and on the way made two inferences.

- Complex programs for different areas have a lot of similarities in design even if those areas are very far from each other.

- Programs for each particular area show no significant changes or improvements throughout the last 15 years. There could be some (even significant) changes in the engine (invisible) part, but nothing really new in interfaces.

When these two statements are put side by side, they highlight the existence of one big problem general for a very vast area: a long stagnation in the whole area of engineering and scientific applications. The problem is in the dominance of developer's view over the use of his product. The developers of applications are usually lesser specialists in each particular area of science or engineering than the researchers and engineers. So the rules of use introduced into the programming products by lesser specialists limit the research work of much better specialists who have to use those applications. But the total control of developers over their products was installed into all the applications from the beginning of computer era; throughout those decades this undeclared rule of full developers' control turned into axiom.

You can't overpower an axiom with a simple declaration like "I think that if you try to do it in a different way it can show better results". You have to demonstrate the advantages of new ideas to developers of programs and to users.

I never considered users of applications as idiots (sorry to mention, but this is a popular view among the developers). The researchers (scientists, engineers) immediately see the advantages of new applications for their work, they stick to these programs, require all the new applications to be developed only of such a type, and they are very much upset that there is no simple method to change in the same way older programs which they continue to use in parallel. I wrote in another recently published article [5] that from time to time I like to watch the scientists' work with new applications. Without any artificial limitations, they can work now with the screen objects as they were doing for years with sheets of paper on which they could draw and write anything. In addition to that old freedom they also get the full power of saving the results, of changing parameters of visualization, of zooming any part in any way, and so on. I would like to demonstrate some example of a scientific application here, but

- The whole idea is based on movability of objects and any picture is static.
- Any specialized application would require a long enough explanation of its purpose.

To avoid such a long and senseless explanation I decided to demonstrate a view of a similar program with easy to understand idea; it's again one more example from [3]. Suppose that you teach a course on the view and features of mathematical functions. You are going to talk about trigonometric functions, exponents, logarithms, and some others, but for any of those functions you would like to demonstrate how coefficients and additional parameters can change the view. At any moment your students can ask you to compare those functions with something new and not predefined. I don't need to go into many other possibilities; **figure 5** shows one view of such program. Certainly, if you would be teaching a course, you would organize the view in a different way; I prepared this view only as a reminder for myself about the set of predefined functions.

In this program user can organize an arbitrary number of plotting areas; each area can show any number of different functions. There is a big group in view; I have already described briefly the use of such groups, so its view can be changed by user at any moment. The group can show the lists of predefined and new functions. To define a new function, you press a button and another form is opened. There you can define a new function by typing it in a standard mathematical notation, and then this application uses its own interpreter. Certainly, everything is movable, resizable, and changeable; everything is saved and restored. It's a classical user-driven application without any limitations.

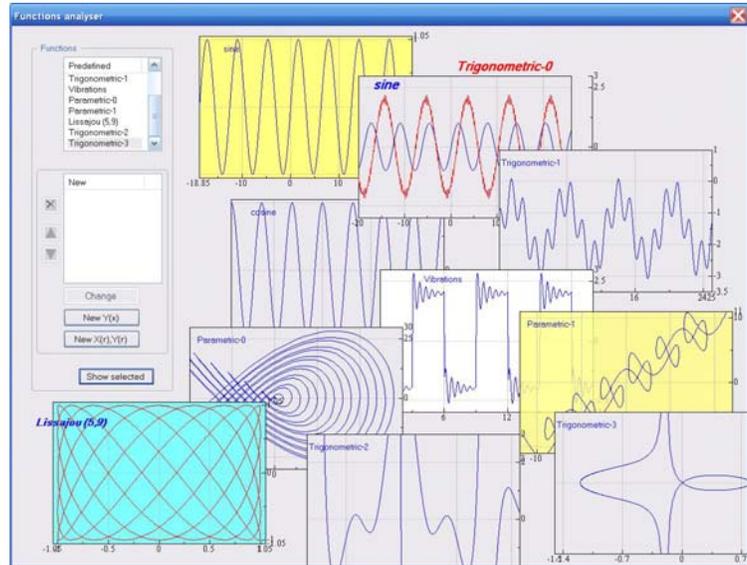

**Fig. 5** An application to demonstrate functions

One of the main features of standard interface, and this is really a nice and valuable feature, that the same controls work identically in all the programs, so users do not need to learn a lot of new things when they start any new application. For user-driven applications the main rule is the movability of each and all elements, so users don't need to read instructions for each new product. If they see a plotting area, they know that it can be moved by any point and resized by borders and corners. Scales and comments can be moved to any location individually, but they retain their positions during the moving / resizing of the plotting area with which they are associated. There can be some very specific requirements from users for one or another application, but this is normal when you design some very specialized program.

The main feature is the passing of the full control from designer to users. For years I was developing applications according to the users' requirements; in the standard version of the same application I would have to discuss the number of possible plotting areas, their positions in each case, and would have to think about not trivial technique to change this number on the fly. For each number of plotting areas I would have to predetermine the best possible configuration and then I would have to develop some mechanism for users to change a view. Not in any way they want but inside the limitations that I would have to impose on them. A classical type of adaptive interface! I was developing such systems 20 years ago. I wouldn't call such development a nightmare; it was really an exciting thing to develop such system. Users even liked it (maybe because they had nothing better to compare with), but in comparison with the scientific applications that I give to users today it was a masterpiece of the Stone Age.

For years the preliminary discussion of a future application between the developers and would be users (scientists, engineers) was not an easy process which was often accompanied by quarrels. All the books on software engineering declare the first rule of design: never start any development of a big program until the specification is thoroughly discussed and every detail is protocoled. If you start without it there will be an infinitive stream of changes which will result in patches over patches and never ending development. Unfortunately, this rule sounds perfectly but can't be used in real life. By the core nature of research work nobody can predict the future outcome in a month or two from today and the requirements to the used applications are going to change in parallel with the research. The second very important obstacle is in different languages that reserchers and programmers prefer to speak. Researcher (engineer) can explain his requirements at the best in the normal language and terms of his professional area; developer immediately tries to translate it into the terms of classes, events, screen objects, and so on. It's difficult to discuss the new projects even with the engineers (scientists) who were writing programs themselves for many years and continue to look at any product from the base of what they learnt 20 and more years ago. It's much better when the researchers explain their requests in their professional terms with which they are proficient and exact. The design of user-driven applications is excellent for such discussion.

For example, the request for an application to teach mathematical functions will be short and accurate in such a form.

- An application must allow to demonstrate a set of predetermined functions and any arbitrary functions of Y(x) type or parametric functions {X(r), Y(r)}.

- Text of any new function can be typed in by using a set of standard functions { sin, cos, tg, sh, ch, th, ln, lg, exp, sqrt, mod, arcsin, arccos, arctg }, arithmetic operations { +, -, *, / }, symbol ^ for degree operation, and brackets.

- Any number of plotting areas can be used for demonstration.

- Any number of functions can be shown in each plotting area.

- Plotting areas are movable and resizable and can be positioned anywhere on the screen.

- The plotting areas can overlap; their order of painting can be easily changed.

- Plotting areas and scales may have an arbitrary number of textual comments. Comments can be added and deleted at any moment; comments can be moved around the screen and placed anywhere; visualization parameters of all the comments must be easily changed.

- All visualization parameters can be changed by user while an application is running.

- Texts of all the functions, all visualization parameters, and all positions of all elements have to be saved on closing an application. The next start of an application must restore the previously saved view.

When such an application is under design, users can start to work with it even from the beginning. It is possible that developer would forget to organize the changing of one or another visualization parameter or would forget to save / restore some of the parameters. Fixing of these bugs will be easy and without any significant effect on the work of an application. Users may also find the lack of some helpful features which were not mentioned in the above shown list; these things would be easy to add without any redesign of an application.

As a developer of such application I am responsible for correct work of the engine (interpreter) and… nothing else. Everything else is under users' control. Don't even think about telling users of such applications how they have to work with them. Now it's their territory.

## Conclusion

I could write an article "On movability of the screen objects" many years ago. That article would begin with mentioning of previous works of different specialists who happened to include words *movable* or *movability* into their publications. There would be a solid list of references and it doesn't matter at all that none of those works has anything to do with the proposed and discussed idea. That article would be not based on any achievements at all. It would be enough to explain the greatness of idea, claim the priority in this area, and add somewhere closer to the end that "now I work on the implementation of this idea in very significant applications; the results will be reported shortly". This would be a classical scientific paper; I saw many of them throughout my life.

Instead, I spent years on the algorithm for turning any object into movable / resizable and on development of the theory of user-driven applications. Only because the design of engineering and scientific applications went into stagnation somewhere 12 – 15 years ago and there seemed to be no end to that stagnation period. Thoughts about the real cause of this stagnation brought me to some conclusions and sparked the work on the algorithm. My long and numerous searches throughout the past results only proved that there were no previous successful attempts to build a system or a big enough application on the basis of movable objects and with a full control passed to users. As one of my colleagues liked to describe similar tasks years ago: "It is impossible because it is never possible."

Yet, it turned out to be possible and produced the outstanding results. The algorithm works now in applications for many different areas; this proves only the universality of this algorithm. But the algorithm itself is not the main result of my work. The idea and implementation of user-driven applications is much more important. You can use my algorithm or design something of your own. I am sure that other algorithms for moving all the screen objects will be produced, but any algorithm will be a foundation for the same type of programs which I call *user-driven*.

The main idea is in passing the full control over applications to users and this can be done only when users can do whatever they want with all the screen objects.

Dr. Sergey Andreyev ( andreyev_sergey@yahoo.com )

September 2011